\documentclass[conference]{IEEEtran}
\IEEEoverridecommandlockouts
\usepackage{cite}
\usepackage{amsmath,amssymb,amsfonts}
\usepackage{algorithmic}
\usepackage{graphicx}
\usepackage{textcomp}
\usepackage{xcolor}
\usepackage{booktabs}
\usepackage{url}
\usepackage{comment}
\usepackage{soul}

\def\BibTeX{{\rm B\kern-.05em{\sc i\kern-.025em b}\kern-.08em
    T\kern-.1667em\lower.7ex\hbox{E}\kern-.125emX}}
\begin{document}

\title{PHONOS: PHOnetic Neutralization for Online Streaming Applications
% {\footnotesize \textsuperscript{*}Note: Sub-titles are not captured for https://ieeexplore.ieee.org  and
}

\author{
\IEEEauthorblockN{Waris Quamer, Mu-Ruei Tseng, Ghady Nasrallah and Ricardo Gutierrez-Osuna}
\IEEEauthorblockA{
\textit{Department of Computer Science and Engineering} \\
\textit{Texas A\&M University}\\
College Station, United States \\
\{quamer.waris, mtseng, ghadynasrallah, rgutier\}@tamu.edu
}
}

% \and
% \and
% \and
% \and
% \and

\maketitle

\begin{abstract}
Speaker anonymization (SA) systems modify timbre while leaving regional or non-native accent cues intact, which is problematic because such cues can reveal a speaker's first-language or geographic background and narrow the anonymity set. To address this issue, we present PHONOS, a streaming module for real-time SA that performs accent neutralization in a privacy sense: reducing accent-origin cues by converting non-native segmental realizations toward a chosen target accent domain. Our approach pre-generates \emph{golden speaker} utterances that preserve source timbre and rhythm but replace foreign segmentals with native ones using silence-aware DTW alignment and zero-shot voice conversion. These utterances supervise a \emph{causal} accent translator that maps non-native content tokens to native equivalents with at most 40ms look-ahead, trained using joint cross-entropy and CTC losses. Our evaluations show an 81\% reduction in non-native accent confidence, with listening-test accentedness ratings consistent with this shift. PHONOS also moves outputs away from the original speaker in embedding space, suggesting lower linkability under an embedding-based proxy, while running with $\leq241\,\mathrm{ms}$ end-to-end latency on a single GPU.
\end{abstract}

\begin{IEEEkeywords}
streaming accent conversion, static traits, voice conversion, privacy preservation, pronunciation correction.
\end{IEEEkeywords}

\section{Introduction}
\label{sec:intro}

Beyond linguistic content, a speech signal encodes a rich set of biometric and paralinguistic cues such as speaker identity, age, gender, emotional state, and accent, any of which can be exploited by adversaries for recognition, profiling, or surveillance \cite{tomashenko2024vpc, patino2021speaker}. As voice-based interfaces proliferate and privacy regulations tighten, protecting these attributes without degrading the communicative utility of speech has become an urgent challenge. The VoicePrivacy Challenge \cite{tomashenko2024vpc}, now in its third edition, has formalized the evaluation of speaker anonymization systems that balance privacy against intelligibility. In parallel, IARPA's Anonymous Real-Time Speech (ARTS) program \cite{iarpa_arts} has pushed the development of streaming solutions that must operate under tight latency budgets while preventing speaker re-identification, modifying static traits such as dialect and age, and suppressing dynamic traits such as emotion.

The dominant approach to speaker anonymization follows a voice-conversion template: decompose speech into content and speaker representations, replace the speaker embedding with a pseudo-speaker identity, and resynthesize \cite{srivastava2020design, meyer2023anonymizing}. Recent streaming systems based on causal CNN-transformer architectures \cite{quamer2024slt24, quamer2025darkstream}, time-varying timbre representations \cite{tvtsyn}, and language-model decoders \cite{cai2025genvc, paniquex2026streamvoiceanon} have shown strong anonymization performance under sub-second latency. Yet these systems share a critical blind spot: they modify \emph{who} the speaker sounds like, but do not explicitly change \emph{where} the speaker sounds like they are from. Accent--a static trait that reflects a speaker's first language, geographic origin, and sociolinguistic background--may remain partially preserved after voice conversion.

This omission matters because accent is a privacy-relevant attribute that affects speaker linkability, personal attribute inference, and social perception. Speaker verification systems exhibit strong accent-dependent bias \cite{ferrer2022fairness}, and both forensic and automated methods exploit accent cues to infer geographic origin and background \cite{rosenhouse2017forensic,peerj2024social_profiling}. As a result, accent has been explicitly identified as personally identifiable information that can be exploited in privacy attacks \cite{Nespoli_2024}. Beyond linkability, accent also shapes social perception in consequential ways: non-native speakers are judged as less trustworthy and competent, with downstream effects on hiring decisions, courtroom credibility, and access to services \cite{levari2010accent,fuertes2012review,baquiran2020doctor,geiger2023accent,frontiers2025forensic_accents}. 

Prior work on phonetic neutralization, commonly referred to as foreign-accent conversion (FAC), has focused on computer-assisted pronunciation training (CAPT) for non-native learners, with early research on "imitation training" suggesting that learners can improve their pronunciation by listening to a voice similar to their own but exhibiting native pronunciations (i.e., a golden speaker) \cite{probst2002enhancing, hirose2003pronunciation}. Thus, the benefits of neutralizing accents goes beyond privacy considerations and can improve communication in a new language\cite{zhao2021converting,ding2019golden,van2014listening,quamer2025disentangling}, making FAC a must-have capability to existing voice anonymization pipelines. Early FAC methods required parallel native references at inference~\cite{aryal2013interspeech, zhao2019interspeech}. Zhao et al.~\cite{zhao2021converting} removed this constraint with a two-stage reference-free approach using golden speaker synthesis followed by pronunciation correction. Subsequent work extended FAC to zero-shot multi-speaker settings via semantic translation with minimal parallel data~\cite{quamer2022zeroshot, jiang2024zeroshot_accent}, and LLM-based multitask frameworks~\cite{cheng2025speechaccentllm}. Most recently, Nguyen et al.~\cite{nguyen2025streaming} proposed a streaming accent conversion model, though with a high latency (0.8\,s). Despite this progress, the literature does not report on FAC systems that can operate with low-latency ($\leq250\,\mathrm{ms}$) for streaming, nor has accent-neutralization been evaluated as a privacy-enhancing component within a speaker anonymization pipeline.

% In this paper, we propose PHONOS, a streaming phonetic neutralization system\footnote{Demo: \url{https://anonymousis23.github.io/demos/phonos}} designed for real-time foreign accent conversion with support for voice anonymization.
% The system adapts the reference-free FAC paradigm of Zhao \textit{et al.}~\cite{zhao2021converting} to operate under causal, low-latency constraints using TVTSyn~\cite{tvtsyn} as the synthesis backbone, achieving accent neutralization with $\leq$120\,ms future context and $\leq$241\,ms end-to-end GPU latency. We evaluate accent reduction using both an accent-classifier probe and human accentedness ratings, and analyze synthesis quality with NISQA-MOS and subjective MOS. Our evaluations show an 81\% reduction in non-native accent confidence with listening-test ratings consistent with this shift, and reduced speaker linkability as accent-neutralized utterances move away from the original speaker (lower embedding cosine similarity).

In this paper, we propose PHONOS\footnote{Demo: \url{https://anonymousis23.github.io/demos/phonos}}, a streaming model for real-time foreign accent conversion and voice anonymization. We use the expression \emph{phonetic neutralization} in a privacy-preservation sense, where the goal is not to produce native-accented speech, but to reduce cues that can reveal a speaker's first-language or linguistic background. PHONOS adapts a reference-free FAC approach to causal, low-latency inference using TVTSyn~\cite{tvtsyn} as the streaming synthesis backbone, achieving reductions in source-accent cues with $\leq120\,\mathrm{ms}$ future context and $\leq241\,\mathrm{ms}$ end-to-end GPU latency.

Our contributions are threefold:
\begin{itemize}
\item We formulate accent neutralization as a privacy-motivated source-accent cue reduction problem for real-time speaker anonymization.
\item We introduce a silence-aware golden-speaker generation pipeline that preserves source-speaker timbre and timing while providing native-accent supervision for streaming training.
\item We propose and evaluate a non-autoregressive causal token-space accent translator that achieves strong source-accent cue reduction under low-latency constraints, with comparisons against reconstruction controls, offline baselines, and golden-speaker upper bounds.
\end{itemize}

% (i) we formulate accent neutralization as a privacy-motivated source-accent cue reduction problem for real-time speaker anonymization, targeting accent-origin cues that remain after conventional timbre-based anonymization;
% (ii) we introduce a silence-aware golden-speaker generation procedure that aligns native-accent content to non-native timing while preserving the source speaker's timbre and pause structure, producing duration-matched supervision for streaming training;
% (iii) we propose a causal, non-autoregressive accent translator that operates in TVTSyn's discrete token space and is trained with joint frame-level cross-entropy and sequence-level CTC objectives; and
% (iv) we evaluate PHONOS on unseen Indian-English speakers using accent-probe, subjective accentedness, quality, speaker-similarity, baseline, and runtime analyses, showing that it approaches offline accent-conversion references while preserving the intended low-latency streaming operating point.}

\section{Method}
\label{sec:method}

PHONOS is a streaming foreign accent conversion (FAC) system for voice anonymization that modifies segmental/phonetic realizations while preserving or anonymizing speaker identity under real-time constraints. We follow a two-stage training paradigm inspired by reference-free FAC \cite{zhao2021converting}, redesigned for causal inference, duration preservation (including pauses), and integration with a streaming speech synthesizer (TVTSyn \cite{tvtsyn}). Stage~I generates duration-matched golden speaker utterances offline, while Stage~II trains a streaming accent translation model that operates fully online.

\subsection{Problem formulation}

Foreign accent conversion aims to modify pronunciation patterns of non-native (L2) speech while preserving linguistic content and speaker identity \cite{zhao2021converting}. In PHONOS, we reformulate this problem as \emph{accent neutralization} for privacy-preserving speech generation. We use the term neutralization not to imply the existence of "accent-free" speech. Specifically, neutralization denotes reducing cues to accent origin that may allow a listener or automated model to infer a speaker's first-language, regional, or sociolinguistic background. Operationally, this corresponds to converting speech from a source-accent domain toward a specified target-accent domain while preserving the intended linguistic content.

Because L2 speakers generally cannot produce native-accented versions of their own utterances, direct ground-truth supervision is unavailable. Prior reference-free FAC work \cite{zhao2021converting} addresses this by generating golden-speaker utterances offline and training a pronunciation correction model to map L2 speech to these generated targets. We adopt this paradigm, but impose additional constraints beyond the original work required for streaming voice anonymization: (i) causal or low-lookahead inference ($\leq$120 ms), (ii) strict duration preservation to support frame-level streaming supervision, and (iii) support for both identity-preserving and identity-anonymizing inference.

In this study, we focus on neutralizing Indian-accented English (INE) (source accent) using General American English (GAE) as the target accent, not to suggest that GAE is universally considered a neutral or accent-free dialect. Under this formulation, PHONOS does not remove accent in an absolute linguistic sense; rather, it reduces INE-specific accent-origin cues by converting segmental/phonetic realizations toward the chosen GAE target domain.

\begin{figure}[t]
    \centering
    \includegraphics[width=0.9\linewidth]{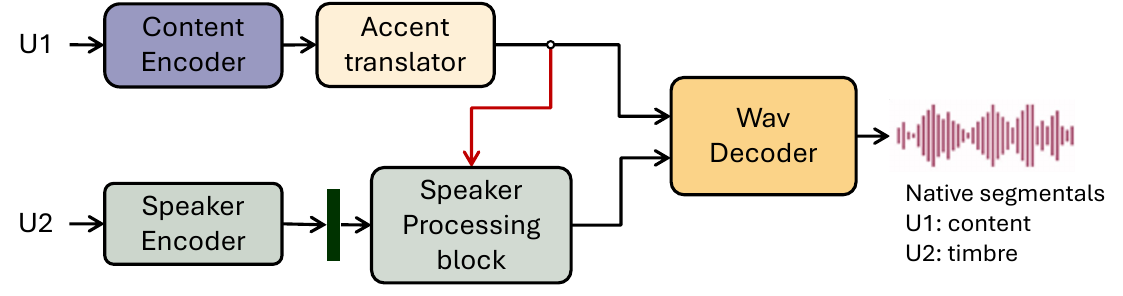}
    \vspace{-10pt}
    \caption{PHONOS inference pipeline. Non-native speech is encoded into content tokens, accent-translated to native tokens, and decoded into a waveform conditioned on the original or pseudo-speaker embedding.}
    \label{fig:inference}
    \vspace{-10pt}
\end{figure}

\begin{figure}[t]
  \centering
  \includegraphics[width=0.9\linewidth]{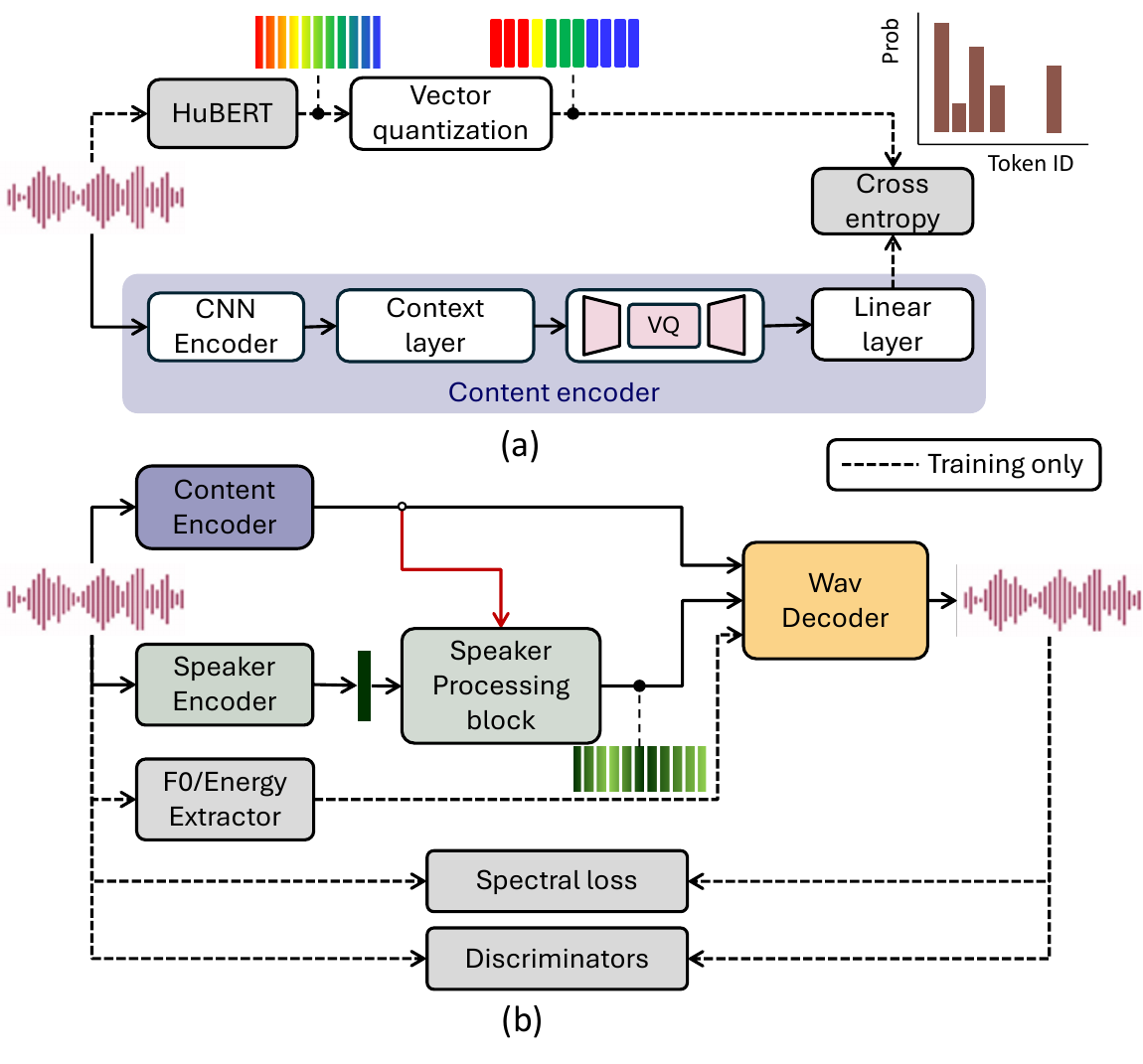}
  \vspace{-15pt}
  \caption{TVTSyn training workflow. (a) content encoder trained against  HuBERT k-means pseudo-labels, and (b) wav decoder conditioned on speaker embedding trained with self-supervision and discriminator objectives.}
  \label{fig:tvtsyn}
  \vspace{-15pt}
\end{figure}

\subsection{System overview}
Formally, given a non-native utterance $x^{\mathrm{L2}}$, we generate a golden speaker utterance $\hat{x}^{\mathrm{GS}}$ that combines native segmental pronunciation (i.e., content tokens) with non-native timbre and timing. A \underline{causal} pronunciation correction model $f_\theta$ is then trained such that $f_\theta(x^{\mathrm{L2}}) \approx \hat{x}^{\mathrm{GS}}$, operating without native references at inference. At inference (Figure~\ref{fig:inference}), the system operates as a streaming pipeline consisting of: (i) a content encoder that extracts discrete, speaker-independent tokens from incoming speech, (ii) an accent translator that maps non-native tokens to native tokens, and (iii) a waveform decoder conditioned on a speaker embedding.

The accent translator can be viewed as performing machine translation in phonetic token space, i.e., mapping between pronunciation domains rather than languages. The content encoder and decoder are inherited from TVTSyn \cite{tvtsyn}, while the accent translator is the key contribution of this work.

\noindent \textbf{TVTSyn backbone.}
Shown in Figure~\ref{fig:tvtsyn}, TVTSyn is a streaming speech synthesizer designed for low-latency voice conversion and anonymization \cite{tvtsyn}. Its content encoder uses a causal CNN followed by transformer layers with a 2\,s look-back window and 80\,ms look-ahead, producing 50\,Hz frame embeddings quantized through a factorized VQ bottleneck and trained with cross-entropy against HuBERT $k$-means pseudo-labels ($N{=}200$). The decoder mirrors this architecture but without any lookahead in the transformer layers and reconstructs waveforms via causal transposed convolutions. TVTSyn achieves as low as 80\,ms GPU latency; we refer readers to~\cite{tvtsyn} for full architectural details. For this study, we made the following changes to the original architecture of TVTSyn, (i) we assigned 40\,ms lookahead to both encoder and decoder blocks, resulting in the same overall lookahead setting as the original implementation, and (ii) fed the decoded tokens rather than the VQ block output of the encoder. In our PHONOS formulation, TVTSyn provides the streaming content encoder and waveform decoder; the contributions specific to PHONOS are the duration-matched golden-speaker construction and the causal token-space accent translator as described below.

\begin{figure}[t]
    \centering
    \includegraphics[width=0.85\linewidth]{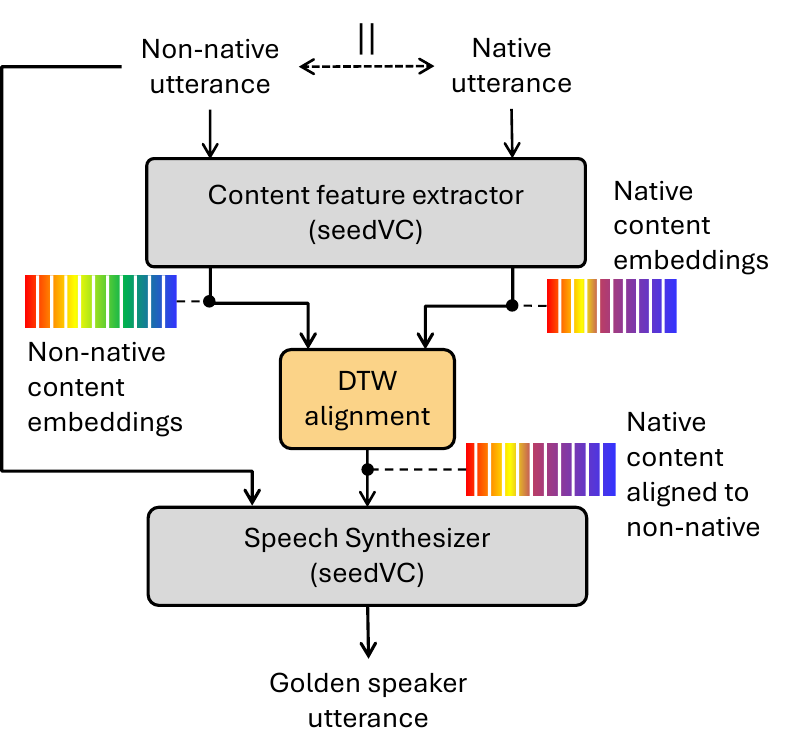}
    \vspace{-15pt}
    \caption{Golden speaker generation. Native and non-native content embeddings are duration-aligned via silence-aware DTW, then synthesized with the non-native speaker's identity.}
    \label{fig:gs_generation}
    \vspace{-20pt}
\end{figure}

\subsection{Stage~I: Golden Speaker generation}
\label{sec:gs_generation}

Golden speaker generation is performed offline to construct duration-matched
supervision for the streaming accent translator. For each L2 utterance, we
synthesize a target-accent L1 utterance from the same transcript using the
TTS system described in Section~\ref{sec:experiments}. Thus, Stage I does not require naturally recorded cross-accent parallel speech. The synthesized L1 utterance and
original L2 utterance form a pseudo-parallel pair from which target-accent segmental content is transferred onto the L2 speaker's timing and voice. Since the two utterances share the same lexical content but differ in accent realization, the resulting golden-speaker utterance provides a supervised target that preserves the L2 speaker's rhythm, pause structure, and timbre while replacing source-accent segmental cues with target-accent realizations. This is critical for Stage~II, where the accent translator is trained with frame-level supervision and must preserve input-output timing during online inference. The procedure consists of two steps: silence-aware alignment and synthesis, as shown in Figure~\ref{fig:gs_generation}.

\noindent \textbf{Silence-aware alignment.}
We extract content features from the L1 and L2 utterances using the Whisper-small encoder from Seed-VC~\cite{liu2024seedvc}. Direct DTW over full utterances often produces poor alignments around pauses and silence boundaries. To avoid this, we first remove silence regions using Silero VAD~\cite{silero_vad}, perform DTW only over non-silent segments, and resample the L1 content sequence to the L2 timing grid. We then reinsert silence at the original L2 positions. This yields an aligned native-accent content sequence with the same duration and pause structure as the L2 utterance.

\noindent \textbf{Synthesis.}\footnote{
Synthetic parallel data released at: \url{https://huggingface.co/datasets/warisqr007/GAPS}
}
The aligned native-accent content sequence is combined with a speaker embedding extracted from the L2 utterance and passed to the Seed-VC synthesizer~\cite{liu2024seedvc}, which uses a flow-matching diffusion transformer with a BigVGAN-v2 vocoder~\cite{lee2022bigvgan}. The resulting golden-speaker waveform combines target-accent pronunciation with the L2 speaker's timbre and timing. These waveforms are used only as offline training targets and are not required during PHONOS inference.

\begin{figure}[t]
    \centering
    \includegraphics[width=\linewidth]{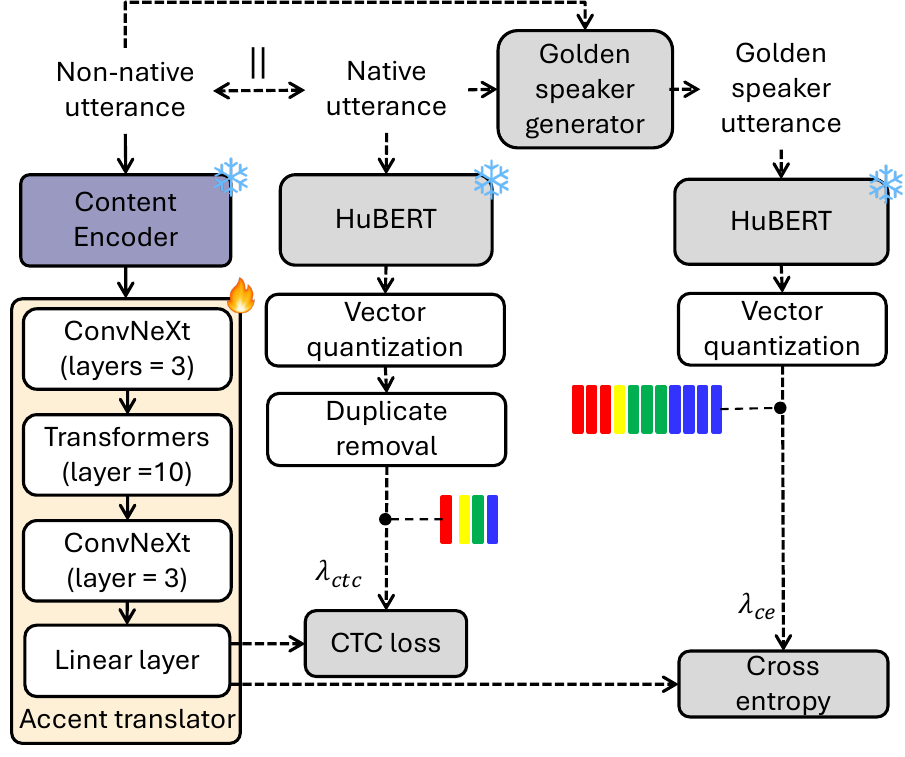}
    \vspace{-20pt}
    \caption{PHONOS's accent translator architecture. Non-native content tokens pass through ConvNeXt and limited-context transformer layers to produce native content tokens.}
    \label{fig:accent_translation}
    \vspace{-15pt}
\end{figure}

\subsection{Stage~II: streaming pronunciation correction}
\label{sec:pronunciation_correction}

We train a non-autoregressive \emph{accent translator} that maps L2 content tokens to L1 content tokens in the TVTSyn discrete token space, using golden speaker utterances as supervision--see Figure~\ref{fig:accent_translation}.

\noindent\textbf{Architecture.}
L2 speech is first encoded into discrete tokens by the frozen TVTSyn content encoder and VQ bottleneck. The accent translator comprises: (i)~a front ConvNeXt block (3 causal layers) for local phonetic patterns, (ii)~a transformer block (10 layers, 8 heads) with 500\,ms past and 40\,ms future context for coarticulatory dependencies, and (iii)~a rear ConvNeXt block (3 causal layers) with a gated skip connection from the transformer output. A linear head projects to $201$ classes (200 content codes + CTC blank).

\noindent\textbf{Training.}
The model is trained with a dual loss $\mathcal{L} = \lambda_{\mathrm{CE}}\,\mathcal{L}_{\mathrm{CE}} + \lambda_{\mathrm{CTC}}\,\mathcal{L}_{\mathrm{CTC}}$. Cross-entropy is computed against frame-aligned HuBERT pseudo-labels from the golden speaker utterances, exploiting the DTW alignment for fine-grained supervision. CTC is computed against de-duplicated native token sequences, providing robustness to residual alignment errors since it requires no frame-level correspondence. We trained ablations for different values for $\lambda_{\mathrm{CE}}$ and $\lambda_{\mathrm{CTC}}$ and found best model performance for $\lambda_{\mathrm{CE}}{=}\lambda_{\mathrm{CTC}}{=}1.0$.

\noindent\textbf{Inference.}
The system operates fully online: L2 speech is processed frame-by-frame through the content encoder, accent translator, and decoder. Accent-translated tokens replace the original L2 tokens before decoding. We use top-$k$ sampling ($k{=}10$, $T{=}0.7$) for natural token diversity.

\begin{table*}[t]
\centering
\caption{Accent neutralization, quality, intelligibility, and speaker similarity evaluation on L2-ARCTIC Indian English speakers. $C_{\mathrm{NA}}$: \% of utterances classified as a native accent. $C_{\mathrm{NN}}$: \% classified as non-native. $\Delta p$: mean posterior probability shift toward native accent classes relative to the original Indian English input (higher is better). Accentedness: subjective foreign accentedness on a 9-point scale (lower is better). NISQA-MOS: predicted MOS for speech quality. MOS: subjective quality on a 5-point scale. WER: word error rate computed against ground-truth transcripts using an automatic ASR system (lower is better). SpkSim: cosine similarity between the output and the GT-INE speaker embedding. Subjective ratings were collected for the ground-truth, TVTSyn, Golden Speaker, and PHONOS conditions.}
\label{tab:accent_probe_indian}
\resizebox{\linewidth}{!}{
    \begin{tabular}{lccccccccc}
    \toprule
    \textbf{Input type} 
    & $\boldsymbol{C_{\mathrm{NA}}}$ \textbf{(\%)} ($\uparrow$) 
    & $\boldsymbol{C_{\mathrm{NN}}}$ \textbf{(\%)} ($\downarrow$) 
    & $\boldsymbol{\Delta p}$ ($\uparrow$) 
    & \textbf{Accentedness} ($\downarrow$) 
    & \textbf{NISQA-MOS} ($\uparrow$) 
    & \textbf{MOS} ($\uparrow$) 
    & \textbf{WER}($\downarrow$)
    & $\boldsymbol{SpkSim}$ \\
    \midrule
    \textbf{GT-GAE} & 100.0 & 0.0 & - & $3.90 \pm 2.07$ & $4.84 \pm 0.20$ & $3.92 \pm 0.89$ & $0.05 \pm 0.09$ & - \\
    \textbf{TVTSyn-GAE} & 100.0 & 0.0 & - & $3.87 \pm 2.09$ & $4.46 \pm 0.40$ & $3.70 \pm 0.89$ & $0.07 \pm 0.14$ & - \\
    \textbf{GT-INE} & 13.5 & 86.5 & - & $5.00 \pm 2.28$ & $3.86 \pm 0.57$ & $3.83 \pm 0.87$ & $0.11 \pm 0.15$ & - \\
    \textbf{TVTSyn-INE} & 57.5 & 42.5 & 0.41  & $4.89 \pm 2.11$ & $3.35 \pm 0.51$ & $3.49 \pm 0.99$ & $0.35 \pm 0.24$ & $0.86 \pm 0.04$ \\
    \textbf{Golden Speaker} & 99.2 & 0.8 & 0.83 & $3.98 \pm 1.84$ & $4.13 \pm 0.52$ & $3.75 \pm 0.86$ & $0.09 \pm 0.14$ & $0.80 \pm 0.05$ \\
    \midrule
    \emph{Offline baselines} & & & & & & & & \\
    \textbf{VEVO} \cite{zhang2025vevocontrollablezeroshotvoice} & 90.3 & 9.7 & 0.78 & - & $4.05 \pm 0.59$ & - & $0.35 \pm 0.24$ & $0.80 \pm 0.07$ \\
    \textbf{SeedVC v2} \cite{liu2024seedvc} & 99.5 & 0.5 & 0.84 & - & $4.62 \pm 0.31$ & - & $0.31 \pm 0.26$ & $0.50 \pm 0.04$ \\
    \midrule
    \emph{Proposed streaming system} & & & & & & & & \\
    \textbf{PHONOS} & 97.0 & 3.0 & 0.81 & $4.14 \pm 1.45$ & $3.33 \pm 0.88$ & $3.49 \pm 0.91$ & $0.24 \pm 0.21$ & $0.72 \pm 0.08$ \\
    \bottomrule
    \end{tabular}
    }
\vspace{-10pt}
\end{table*}

\section{Experimental setup}
\label{sec:experiments}

\noindent\textbf{Training corpus for accent conversion.}
TVTSyn is trained on LibriTTS \cite{zen2019librittscorpusderivedlibrispeech} corpus. We train our accent translator module using original-golden speaker pairs derived from the IndicTTS\footnote{\url{https://www.iitm.ac.in/donlab/indictts/database}} \cite{baby2016resources} corpus, which contains approximately 250 hours of Indian-accented English speech from 25 speakers. All recordings are sampled at 16\,kHz.

\noindent\textbf{Native speech corpus for Golden Speaker generation.}
Native English speech for golden speaker generation is generated using an off-the-self TTS system (F5-TTS \cite{chen-etal-2024-f5tts})
% \footnote{\url{https://github.com/SWivid/F5-TTS}} ) 
using the available ground truth transcription from IndicTTS corpus. Native utterances are used only during the offline golden speaker generation stage and are discarded afterwards.

\noindent\textbf{Evaluation.}
We evaluate our system on the Indian subset of L2-ARCTIC \cite{zhao2018l2}, so the evaluation is cross-corpus and on unseen speakers. 
The objective evaluation uses all four Indian-English speakers in L2-ARCTIC,
with 100 randomly sampled utterances per speaker (400 utterances total). We perform synthesis in self-reconstruction mode, \textit{i.e.}, we condition the system on the speaker embedding extracted from the same source utterance that is fed to the content encoder. For each subjective test, 20 listeners rated 90 samples, with 15 samples
per evaluated condition.
% Because there is no standard benchmark and no available open-source foreign accent conversion baseline, we do not report direct baseline comparisons. Instead, we treat the golden-speaker utterances as an upper bound and use them as our reference baseline.

Because no prior system matches our complete operating point of reference-free, low-latency streaming FAC integrated with a streaming anonymization backbone, we organize comparisons into three groups. First, we include reconstruction controls using the TVTSyn backbone without accent translation: we use TVTSyn to reconstruct INE utterances (TVTSyn-INE) and GAE utterances (TVTSyn-GAE). Second, we include Golden Speaker utterances as an offline upper bound, since these are the Stage-I supervision targets generated using aligned native pronunciation and the source speaker identity. Third, we compare against VEVO \cite{zhang2025vevocontrollablezeroshotvoice} and SeedVC v2\cite{liu2024seedvc} as \underline{non-streaming / offline} zero-shot voice conversion reference systems that support accent conversion. Note that VEVO and SeedVC v2 are not latency-matched baselines; instead, using them for comparison helps us contextualize how much accent reduction, quality change, and speaker-embedding displacement can be achieved when the causal streaming constraint is relaxed. Although Stage~I target construction also uses the SeedVC family of models, PHONOS does not use SeedVC at inference time.

\noindent\textbf{Model training.} All modules were trained using the AdamW optimizer with an initial learning rate of \(5 \times 10^{-4}\) and a batch size of 16 and an \texttt{ExponentialLR} scheduler with decay factor \(\gamma = 0.999996\). The accent translator module was trained for \(250k\) steps on an NVIDIA RTX 3090 GPU. The accent translator introduces $34.5M$ parameters; combined with the TVTSyn backbone, the full PHONOS pipeline totals $125.8M$ parameters.

\section{Experiments and Results}
\label{sec:results}

We conduct objective and subjective experiments to evaluate (i) streaming accent neutralization, (ii) synthesis quality and intelligibility, and (iii) how accent conversion affects speaker verifiability. We compare eight evaluation conditions: \textbf{GT-GAE} (native General American English, upper bound), \textbf{TVTSyn-GAE} (GAE reconstructed through TVTSyn, isolating synthesis artifacts), \textbf{GT-INE} (Indian English baseline), \textbf{TVTSyn-INE} (INE reconstructed through TVTSyn), \textbf{Golden Speaker} (offline Stage~I targets), outputs from VEVO and SeedVC v2 baselines, and \textbf{PHONOS (Proposed)} (full streaming pipeline).

\subsection{Accent Neutralization}
\label{sec:accent_results}

We use an off-the-shelf English accent classifier\footnote{\url{https://huggingface.co/Jzuluaga/accent-id-commonaccent_xlsr-en-english}} as a perceptual probe. Given an utterance $x$, it outputs $p(\cdot \mid x)$ over 16 accents (8 native, 8 non-native). We report \emph{native} and \emph{non-native} classification rates ($C_{\mathrm{NA}}$, $C_{\mathrm{NN}}$) as \% of utterances classified as native and non-native respectively, and a softer confidence metric: given native posterior mass as $p_{\mathrm{NA}}(x)=\sum_{a\in\mathcal{A}_{\mathrm{NA}}}p(a\mid x)$, where $\mathcal{A}_{\mathrm{NA}}$ represents set of native accent classes, we define mean probability shift towards native accent classes as $\Delta p=\frac{1}{N}\sum_i(p_{\mathrm{NA}}(x'_i)-p_{\mathrm{NA}}(x_i))$, here $x'_i$ and $x_i$ are synthesized and original utterances respectively.

Table~\ref{tab:accent_probe_indian} shows the probe cleanly separates original Indian and American speech (GT-INE: $C_{\mathrm{NA}}{=}13.5\%$ vs.\ GT-GAE: $100\%$). Notably, reconstructing Indian English through TVTSyn already increases $C_{\mathrm{NA}}$ to $57.5\%$ ($\Delta p{=}0.41$), suggesting the native-trained discrete content bottleneck partially normalizes non-native realizations by mapping them to nearby native codebook entries, consistent with observations in discrete token pipelines \cite{onda2024pilot}. Stage~I golden speakers reach near-native performance ($C_{\mathrm{NA}}{=}99.2\%$, $\Delta p{=}0.83$), validating the target construction. PHONOS approaches this upper bound ($C_{\mathrm{NA}}{=}97.0\%$, $\Delta p{=}0.81$), with only $3\%$ still classified as non-native; the remaining gap is plausibly due to streaming constraints (40\,ms look-ahead) limiting resolution of accent-bearing coarticulatory transitions.

Compared with the non-streaming reference systems, PHONOS performs favorably despite operating under causal streaming constraints. VEVO improves native-accent classification substantially ($C_{\mathrm{NA}}{=}90.3\%$, $\Delta p{=}0.78$), but remains below that of PHONOS ($C_{\mathrm{NA}}{=}97.0\%$, $\Delta p{=}0.81$), suggesting that the proposed token-space accent translator trained on supervised golden speaker utterances is more effective at shifting Indian-accented English toward the GAE target-accent domain than this self-supervised VC baseline. SeedVC v2 reaches slightly higher accent-probe performance ($C_{\mathrm{NA}}{=}99.5\%$, $\Delta p{=}0.84$), comparable to the Golden Speaker upper bound. However, SeedVC v2 is an offline system and is not constrained by causal inference or limited look-ahead. Thus, PHONOS achieves accent-probe performance close to the strongest offline reference while maintaining the low-latency streaming operating point targeted in this work.

Subjective accentedness ratings (N=20; 9-point Likert, lower is better) corroborate these trends. GT-INE is rated most accented ($5.00 \pm 2.28$), while PHONOS reduces accentedness to $4.14 \pm 1.45$ ($\approx$17\% relative), close to golden speaker ($3.98 \pm 1.84$) and native speech (GT-GAE: $3.90 \pm 2.07$). TVTSyn-INE shows only marginal perceptual improvement ($4.89 \pm 2.11$) aligning with its partial probe shift; PHONOS improvement of accentedness rating over TVTSyn-INE is statistically significant ($p=0.024$). Thus, the listener ratings independently support an accent-reduction effect beyond the shift induced by TVTSyn reconstruction alone.

\subsection{Synthesis Quality and Intelligibility}
\label{sec:quality_results}

We assess objective quality using NISQA-MOS \cite{mittag2021nisqa}. Native speech scores highest (GT-GAE: $4.84 \pm 0.20$), while TVTSyn reconstruction introduces expected synthesis artifacts (TVTSyn-GAE: $4.46 \pm 0.40$). For Indian English, GT-INE is lower ($3.86 \pm 0.57$), likely reflecting accent mismatch to NISQA’s training distribution, and TVTSyn-INE drops further ($3.35\pm0.51$). PHONOS matches TVTSyn-INE ($3.33 \pm 0.88$), indicating accent translation does not add quality degradation beyond the TVTSyn backbone. Golden speakers score higher ($4.13\pm0.52$), consistent with Seed-VC being an offline, high-capacity diffusion system.

While VEVO and SeedVC v2 obtain higher objective quality than PHONOS, this gap is expected because both reference systems operate offline and use higher-capacity generation pipelines, whereas PHONOS is constrained by the streaming TVTSyn backbone. Importantly, PHONOS remains essentially matched to TVTSyn-INE, indicating that the accent translator itself does not introduce measurable additional quality degradation beyond that due to causal inference. Thus, while offline systems provide an upper quality reference, the relevant streaming comparison is between PHONOS and TVTSyn-INE.

Subjective MOS (N=20; 5-point, higher is better) mirrors the objective trends: PHONOS and TVTSyn-INE are comparable ($3.49 \pm 0.91$ vs.\ $3.49 \pm 0.99$), while golden speakers are higher ($3.75 \pm 0.86$) and GT-GAE remains best ($3.92 \pm 0.89$). Overall, the quality cost is dominated by the streaming synthesis backbone rather than the accent translator.

To further assess lexical intelligibility, we compute word error rate (WER) by transcribing each condition with Whisper-base ASR\footnote{\url{https://huggingface.co/openai/whisper-base}} and comparing the output against the ground-truth transcript. For native-accented speech, GT-GAE and TVTSyn-GAE remain comparable ($0.05 \pm 0.09$ vs. $0.07 \pm 0.14$), indicating that the TVTSyn reconstruction pipeline largely preserves lexical content when the input is close to the native-accent domain used to train the content encoder. 
In contrast, Indian-accented speech shows substantially larger degradation after reconstruction: WER increases from $0.11\pm0.15$ for GT-INE to $0.35\pm0.24$ for TVTSyn-INE. This is consistent with a domain mismatch as the TVTSyn content encoder was trained primarily on native English speech and is therefore less robust to non-native phonetic realizations. PHONOS reduces WER to $0.24\pm0.21$, corresponding to an approximately 31\% relative improvement over TVTSyn-INE and outperforming the offline VEVO ($0.35 \pm 0.24$) and SeedVC v2 ($0.31 \pm 0.26$) references.
Nevertheless, PHONOS remains worse than the original GT-INE speech ($0.24$ vs.\ $0.11$ WER), indicating that accent translation recovers only part of the intelligibility loss introduced by the native-trained streaming backbone. This residual gap represents a utility trade-off and motivates future work on more accent-robust streaming content representations.

% So, the token-space accent translator does not merely shift accent classification, but also improves ASR performance relative to reconstructing Indian-accented speech through the native-trained streaming backbone. The Golden Speaker upper bound remains strongest among converted conditions ($0.09 \pm 0.14$), as expected from its offline target construction.

% \sethlcolor{pink}
% \hl{NOTE: If you have space and bandwidth, it would be good to add a t-sne figure showing speaker embeddings before and accent translation, color coded by accents}

\subsection{Speaker Similarity}
\label{sec:spk_results}

To quantify speaker similarity, we compute cosine similarity ($SpkSim$ column in Table~\ref{tab:accent_probe_indian}) between each processed utterance's speaker embedding and the corresponding GT-INE speaker embedding. We interpret this as a linkability proxy rather than a full attacker-based privacy evaluation. Accordingly, these SpkSim results establish embedding-space displacement under the selected speaker encoder, but should not be interpreted as evidence of reduced linkability against an attacker-based ASV system.

TVTSyn-INE preserves high similarity ($0.86\pm0.04$), showing that reconstruction through the streaming backbone largely retains speaker identity cues when accent is not explicitly modified. Golden Speaker and VEVO both reduce similarity to $0.80$, likely reflecting a combination of pronunciation change and synthesis-induced timbre shift. PHONOS reduces similarity further to $0.72\pm0.08$ despite using the same target speaker embedding and streaming backbone as TVTSyn-INE, suggesting that accent conversion changes phonetic and co-articulatory patterns captured by speaker encoders. SeedVC v2 produces the lowest similarity ($0.50\pm0.04$), but this much larger displacement likely reflects stronger timbre or speaker-representation modification rather than controlled accent conversion alone. Thus, PHONOS provides additional embedding-space displacement relative to reconstruction and VEVO, while avoiding the much larger identity shift observed for SeedVC v2.

\subsection{Runtime and streaming performance}

We evaluate real-time and streaming latency by running end-to-end inference in a chunked setting and reporting real-time factor (RTF) and algorithmic latency. 
The reported latency includes input chunk accumulation, required future context, and processing time for the content encoder, accent translator, and decoder.
On a NVIDIA RTX 3090 GPU, with an 80 ms chunk size, the system achieves an RTF of 0.6 and an end-to-end latency of 241 ms, which includes the 80 ms input chunk accumulation and the 120 ms lookahead required by the system. On CPU, the system runs in real time at a larger chunk size: using 160 ms chunks, it achieves real-time throughput with a total latency of 370 ms, again including the 160 ms chunking delay and 120 ms lookahead.

\section{Discussion}
\label{sec:discussion}
We proposed PHONOS, a streaming accent neutralization system that reduces source-accent cues under tight latency budgets without measurable quality degradation beyond the streaming synthesis backbone. In our experiments, PHONOS also lowers speaker-embedding similarity relative to a TVTSyn reconstruction baseline that shares the same backbone and speaker embedding but does not modify accent, indicating that accent contributes identity-relevant phonetic and coarticulatory information. One of the reasons could be that changing segmental realizations perturbs the phonetic patterns and associated spectral structure that speaker encoders rely on, this additional displacement is consistent with lower speaker similarity under the embedding-space proxy used here. These effects suggest that accent neutralization can serve as a complementary component to identity replacement in real-time voice anonymization pipelines. However, we do not evaluate an adaptive or attacker-based ASV system, so these results should not be interpreted as establishing reduced ASV linkability or end-to-end speaker privacy.

Our results complement prior reference-free FAC work~\cite{zhao2021converting,quamer2022zeroshot}, which did not consider streaming constraints, and extend streaming anonymization work~\cite{quamer2025darkstream,tvtsyn}, which primarily focuses on identity replacement. The current evaluation is limited to Indian-accented English. However, the proposed framework is not specific to the INE-to-GAE setting. So, given source-accent speech and a target-accent domain for golden-speaker construction, the same silence-aware alignment and token-space translation process can be extended to other accent pairs. Stage~I targets are constructed using synthesized rather than naturally recorded
target-accent speech, which may introduce TTS-specific characteristics into the supervision. Future work will expand the proposed methodology to additional L1 backgrounds and larger speaker populations, combine with pseudo-identity replacement and evaluate against attacker models using ASV-based linkability metrics, and incorporate prosodic modification to address residual accent cues not captured by segmental conversion.

% In summary, the accent neutralization is a practical, low-latency capability for reducing source-accent cues and speaker-embedding similarity, making it a natural complement to identity replacement in real-time voice anonymization.

\section*{Acknowledgment}
% To be updated after peer-review.
Supported by the Intelligence Advanced Research Projects Activity (IARPA) via Department of Interior/Interior Business Center (DOI/IBC) contract number 140D0424C0066. The U.S. Government is authorized to reproduce and distribute reprints for Governmental purposes notwithstanding any copyright annotation thereon. Disclaimer: The views and conclusions contained herein are those of the authors and should not be interpreted as necessarily representing the official policies or endorsements, either expressed or implied, of IARPA, DOI/IBC, or the U.S. Government.

% \section*{Generative AI Use Disclosure}
Large language models were only used to improve writing clarity, grammar, and style. No experimental design, data analysis, or result interpretation relied on automated tools.

\bibliographystyle{IEEEtran}
\bibliography{mybib}
\end{document}